\documentclass[twocolumn,twoside]{article}
\makeatletter\if@twocolumn\PassOptionsToPackage{switch}{lineno}\else\fi\makeatother

\usepackage{amsmath,amssymb,tabulary,graphicx,times,caption,fancyhdr,alphalph}
\usepackage[utf8]{inputenc}
\usepackage[T1]{fontenc}
\usepackage[numbers,sort&compress]{natbib}
\usepackage{subfig}
\usepackage{abstract}
\usepackage[paperheight=11.66in,paperwidth=8.91in,margin=2cm,headsep=.5cm,top=2.5cm]{geometry}
\renewenvironment{onecolabstract} {\vspace*{-1pc}\trivlist\item[]\leftskip3pc\textbf{\abstractname :}}{\par\noindent\endtrivlist}
 \date{}\DeclareCaptionLabelFormat{figbf}{Figure \textbf{#2}}
\DeclareCaptionLabelFormat{tabbf}{Table \textbf{#2}}
\usepackage{url,multirow,morefloats,floatflt,cancel,tfrupee}
\makeatletter

\AtBeginDocument{\@ifpackageloaded{textcomp}{}{\usepackage{textcomp}}}
\makeatother
\usepackage{colortbl}
\usepackage{soul}
\usepackage{xcolor}
\usepackage{pifont}
\usepackage[nointegrals]{wasysym}
\makeatletter

\def\mcWidth#1{\csname TY@F#1\endcsname+\tabcolsep}

\def\cAlignHack{\rightskip\@flushglue\leftskip\@flushglue\parindent\z@\parfillskip\z@skip}
\def\rAlignHack{\rightskip\z@skip\leftskip\@flushglue \parindent\z@\parfillskip\z@skip}

\@ifundefined{etal}{}{}

\usepackage{ifxetex}
\ifxetex\else\if@twocolumn\@ifpackageloaded{stfloats}{}{\usepackage{dblfloatfix}}\fi\fi

\AtBeginDocument{
\expandafter\ifx\csname eqalign\endcsname\relax
\def\eqalign#1{\null\vcenter{\def\\{\cr}\openup\jot\m@th
  \ialign{\strut$\displaystyle{##}$\hfil&$\displaystyle{{}##}$\hfil
      \crcr#1\crcr}}\,}
\fi
}

\AtBeginDocument{%
  \@ifpackageloaded{endfloat}%
   {\renewcommand\efloat@iwrite[1]{\immediate\expandafter\protected@write\csname efloat@post#1\endcsname{}}}{\newif\ifefloat@tables}%
}%

\def\BreakURLText#1{\@tfor\brk@tempa:=#1\do{\brk@tempa\hskip0pt}}
\let\lt=<
\let\gt=>
\def\processVert{\ifmmode|\else\textbar\fi}

\@ifundefined{subparagraph}{
\def\subparagraph{\@startsection{paragraph}{5}{2\parindent}{0ex plus 0.1ex minus 0.1ex}%
{0ex}{\normalfont\small\itshape}}%
}{}

\newcommand\role[1]{\unskip}
\newcommand\aucollab[1]{\unskip}
  
\@ifundefined{tsGraphicsScaleX}{\gdef\tsGraphicsScaleX{1}}{}
\@ifundefined{tsGraphicsScaleY}{\gdef\tsGraphicsScaleY{.9}}{}
\def\checkGraphicsWidth{\ifdim\Gin@nat@width>\linewidth
	\tsGraphicsScaleX\linewidth\else\Gin@nat@width\fi}

\def\checkGraphicsHeight{\ifdim\Gin@nat@height>.9\textheight
	\tsGraphicsScaleY\textheight\else\Gin@nat@height\fi}

\def\fixFloatSize#1{}
\let\ts@includegraphics\includegraphics

\def\inlinegraphic[#1]#2{{\edef\@tempa{#1}\edef\baseline@shift{\ifx\@tempa\@empty0\else#1\fi}\edef\tempZ{\the\numexpr(\numexpr(\baseline@shift*\f@size/100))}\protect\raisebox{\tempZ pt}{\ts@includegraphics{#2}}}}

\AtBeginDocument{\def\includegraphics{\@ifnextchar[{\ts@includegraphics}{\ts@includegraphics[width=\checkGraphicsWidth,height=\checkGraphicsHeight,keepaspectratio]}}}

\DeclareMathAlphabet{\mathpzc}{OT1}{pzc}{m}{it}

\def\URL#1#2{\@ifundefined{href}{#2}{\href{#1}{#2}}}

\def\UrlOrds{\do\*\do\-\do\~\do\'\do\"\do\-}%
\g@addto@macro{\UrlBreaks}{\UrlOrds}

\edef\fntEncoding{\f@encoding}

\makeatother

\newif\ifmultipleabstract\multipleabstractfalse%
\usepackage[explicit]{titlesec}
\usepackage[overload]{textcase}
\def\NormalBaseline{\def\baselinestretch{1.1}}
 \titleformat{\section}[hang]{\NormalBaseline\filright\large\bfseries}
{\large\thesection.}
{3pt}
{\MakeTextUppercase{#1}}
[]
\titleformat{\subsection}[hang]{\NormalBaseline\filright\bfseries\boldmath}
{\thesubsection.}
{3pt}
{#1}
[]
\titleformat{\subsubsection}[hang]{\NormalBaseline\filright\bfseries\boldmath\itshape}
{\itshape\thesubsubsection.}
{3pt}
{#1}
[]
\titleformat{\paragraph}[runin]{\NormalBaseline\filright\bfseries\boldmath}
{\itshape\theparagraph.}
{3pt}
{#1}
[]
\titleformat{\subparagraph}[runin]{\NormalBaseline\filright\bfseries\boldmath\itshape}
{\itshape\thesubparagraph.}
{3pt}
{#1}
[]

\titlespacing{\section}{0pt}{1.5\baselineskip}{.2\baselineskip}  
\titlespacing{\subsection}{0pt}{1.5\baselineskip}{.2\baselineskip}  
\titlespacing{\subsubsection}{0pt}{1.5\baselineskip}{.2\baselineskip}  
\titlespacing{\paragraph}{0pt}{.5\baselineskip}{10pt}  
\titlespacing{\subparagraph}{0pt}{.5\baselineskip}{10pt}  

\titlespacing\section{0pt}{12pt plus 4pt minus 2pt}{0pt plus 2pt minus 2pt}
\titlespacing\subsection{0pt}{12pt plus 4pt minus 2pt}{0pt plus 2pt minus 2pt}
\titlespacing\subsubsection{0pt}{12pt plus 4pt minus 2pt}{0pt plus 2pt minus 2pt}

\makeatletter\def\oupIndent{1pt}
\def\author#1{\gdef\@author{\hskip-\dimexpr(\tabcolsep)\hskip\oupIndent\parbox{\dimexpr\textwidth-\oupIndent}{#1}}}
\def\title#1{\gdef\@title{\raggedright\bfseries\ifx\@articleType\@empty\else\@articleType\\\fi#1}}
\let\@articleType\@empty \def\articletype#1{\gdef\@articleType{{\normalfont\itshape#1}}}
\fancypagestyle{headings}{\fancyhf{}\fancyhead[CO]{\textit \RunningHead}\fancyhead[CE]{\RunningAuthor}\fancyfoot[R]{\thepage}}

\newcommand\keywords[1]%
  {\begin{flushleft}
   \let\and\\%
   \textbf{Keywords:}\\
   #1
   \end{flushleft}%
  }

\begin{document}


\title{Application of Monte Carlo algorithms to cardiac imaging reconstruction}
\author{J. Zhou\textsuperscript{\alphalph{1}}, A. G. Leja\textsuperscript{\alphalph{1}}, M. Salvatori \textsuperscript{\alphalph{2}}, D. Della Latta \textsuperscript{\alphalph{2}},
            A. Di Fulvio\textsuperscript{\alphalph{1}}
            ~\\[-3pt]\normalsize\normalfont \\
~\textsuperscript{\textit{\alphalph{1}}}
{\textit{Department of Nuclear, Plasma, and Radiological Engineering, University of Illinois at Urbana-Champaign, Urbana, IL 61801, United States\unskip}}\\
~\textsuperscript{\textit{\alphalph{2}}}
{\textit{Fondazione Toscana G. Monasterio, Massa, MS 54100, Italy\unskip}}
}

\def\RunningHead{{Application of Monte Carlo algorithms to cardiac imaging reconstruction}}\def\RunningAuthor{\textit{Zhou \MakeLowercase{\textit{et al.}} }}


\twocolumn[ \maketitle {\begin{onecolabstract}
Monte Carlo algorithms have a growing impact on nuclear medicine reconstruction processes. One of the main limitations of myocardial perfusion imaging (MPI) is the effective mitigation of the scattering component, which is particularly challenging in Single Photon Emission Computed Tomography (SPECT). In SPECT, no timing information can be retrieved to locate the primary source photons. Monte Carlo methods allow an event-by-event simulation of the scattering kinematics, which can be incorporated into a model of the imaging system response. This approach was adopted since the late Nineties by several authors, and recently took advantage of the increased computational power made available by high-performance CPUs and GPUs. These recent developments enable a fast image reconstruction with an improved image quality, compared to deterministic approaches. Deterministic approaches are based on energy-windowing of the detector response, and on the cumulative estimate and subtraction of the scattering component. In this paper, we review the main strategies and algorithms to correct for the scattering effect in SPECT and focus on Monte Carlo developments, which nowadays allow the three-dimensional reconstruction of SPECT cardiac images in a few seconds. 
\end{onecolabstract}
\keywords{Monte Carlo algorithms, SPECT, MPI, Cardiac imaging reconstruction, Scattering correction}}]

\section{Introduction}

Coronary artery disease (CAD) is the leading causes of morbidity and mortality in the United States \cite{cdcStat2018} and one of leading causes worldwide. Myocardial perfusion imaging is a non-invasive imaging modality that provides quantitative blood perfusion information, helps to assess the overall function of the miocardium and to
diagnose CAD symptoms. MPI is based on two main techniques:  single photon emission computed tomography (SPECT) and positron emission tomography (PET). Non-invasive imaging can be followed by coronary angiography to obtain further anatomical imaging, if deemed necessary after MPI.
SPECT has been traditionally the most widely used MPI modality, mainly because of the availability of the used tracers, i.e., technetium-99m and thallium-201. However, SPECT spatial resolution of 12-15 mm is generally poorer compared to the one achieved in PET, typically 4–7 mm. SPECT exhibits also a slower temporal resolution than PET, which does not allow for absolute quantification of perfusion by accurately tracking the activity of the tracer in the arteries and myocardium as a function of time. Nonetheless, recent developments in SPECT detection technology \cite{Kim2006465}, a more accurate application of photon attenuation correction (AC), and advanced image reconstruction algorithms may improve the overall SPECT image quality and retain the benefit of no needing of a cyclotron to produce radiotracers used in PET.
In this paper, we focus on the use of Monte Carlo algorithms to improve SPECT image reconstruction. In the second section, we introduce stochastic computational methods applied to ionizing radiation transport, with reference to the software most used by the scientific community. In the third section, after a brief review of SPECT basic principles, we focus on the application of computational methods to the image reconstruction and artifact mitigation. Deterministic methods are reviewed before focusing on the Monte Carlo methods for scattering mitigation.

\section{Monte Carlo Algorithms for the simulation of radiation transport }

Shortly after computers were invented, their potential in simulating random processes became clear \cite{metropolis}. This simulation process was named "Monte Carlo" after the iconic gambling house at Montecarlo. The earliest developments of Monte Carlo codes to simulate radiation transport phenomena are due to Robert Oppenheimer and colleagues within the Manhattan project, in the Forties. For over twenty years, the use of Monte Carlo simulations for radiation transport applications was mostly limited to the fields of nuclear physics and technology. Once Monte Carlo codes became available to the broader scientific community, their potential use in radiation protection and shielding became relevant. 
Monte Carlo algorithms have the capability of simulating
the radiation transport and recording some features of the radiation field, so called tallies, related to the average behavior of individual particles. The user typically defines the geometry of the problem, and can select the tallies of interest. Monte Carlo can be used to duplicate a statistical process, such as the interaction of ionizing radiation with matter, and is particularly useful for complex problems that cannot be modeled by computer codes based on deterministic methods. The single interactions
in a radiation transport process are simulated sequentially.
The user-selected tallies, and the standard deviation associated with them, are obtained by statistically sampling the probability distributions that govern the radiation transport. The Monte Carlo process is actually implemented by following each particle produced by a source throughout its
life to its death, e.g., absorption, scattering, escape from the problem volume. Probability distributions are
randomly sampled using transport data, such as reaction cross sections or physical models, to determine the outcome at each step of the particle's life. The execution of Monte Carlo algorithms needs extensive computational resources. Performing a Monte Carlo simulation for a given number of source particles is typically much slower compared to performing an experiment to tally the same quantity, and involving the same number of source particles.   MCNP \cite{mcnp}, GEANT4\cite{geant}, Fluka \cite{fluka}, PHITS \cite{phits} and EGS \cite{egs} are general-purpose Monte Carlo codes used for radiation transport.



    
\section{SPECT imaging and scattering effect}
\begin{figure}[htbp!]
\centering
\includegraphics[height = 6cm]{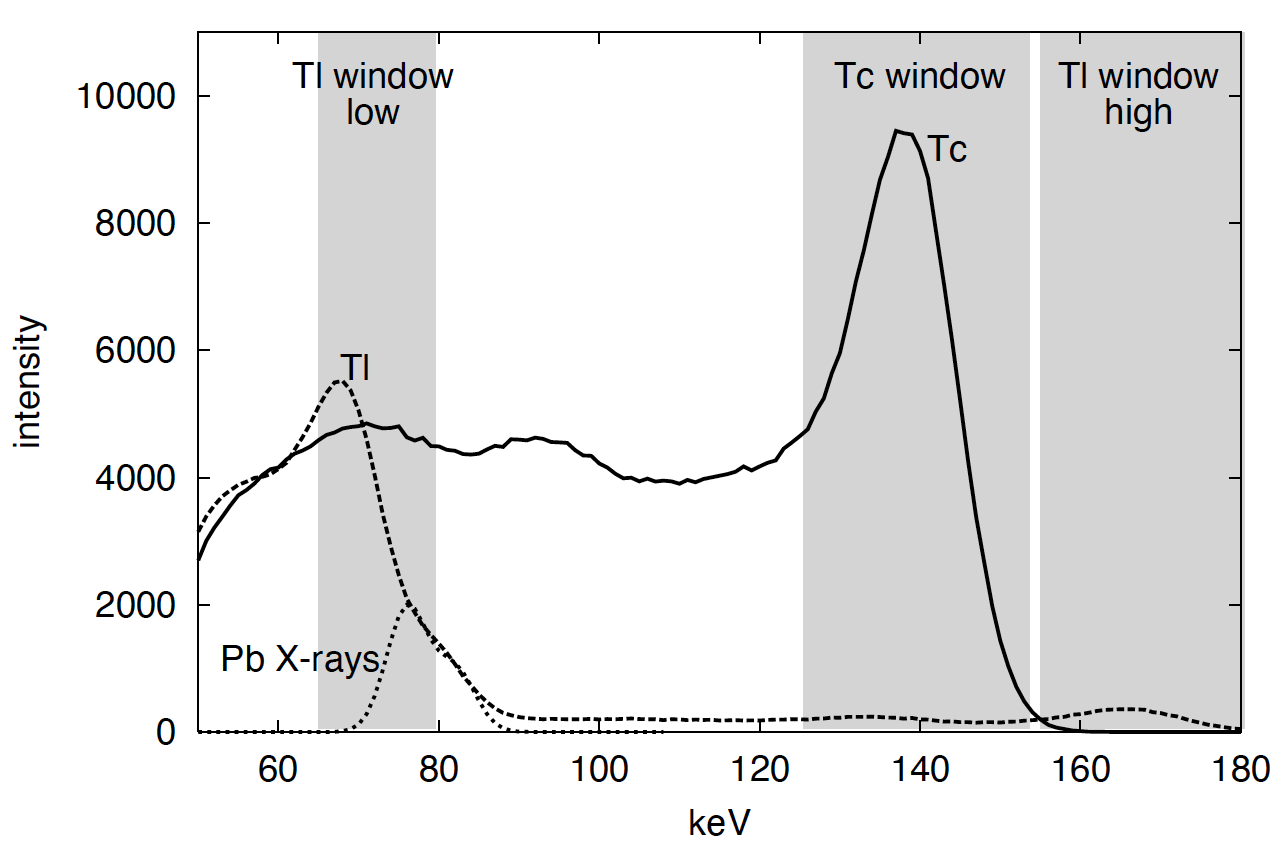}
\caption{Energy spectrum of Tc-99m and Tl-201 in scintillation detector based dual-isotope SPECT. \cite{Jong2001}}
\label{f:Tc_Tl_spectrum}
\end{figure}

SPECT is a nuclear medicine diagnostic procedure, which provides functional three-dimensional information. It is wildly used for functional myocardial perfusion imaging \cite{Matsunari1998,Toyama2004} and brain \cite{Lee1987,Mcdaniel1991} imaging. SPECT uses radioactive tracers emitting photons as information carriers. Radioactive tracers are compounds that can participate in organisms' physiological processes in the same way as non-radioactive molecules, while being detected through their radiation signatures. Once the subject is injected with radioactive tracers, Anger gamma cameras detect the emitted photons and acquire projection data at different angles around the patient. The subsequent reconstruction of the radioactive source attenuation profiles allows to obtain three-dimensional images. Tc-99m and Tl-201 are two most commonly used tracers for perfusion imaging in SPECT. The energy spectrum of two isotopes, acquired during a simulated SPECT procedure, is shown by H. W. Jong \cite{Jong2001}, Figure \ref{f:Tc_Tl_spectrum}. 
Tc-99m emits 140 keV gamma rays, while Tl-201 emits 71 keV (low energy) and 167 keV (high energy) gamma rays. The spectroscopic acquisition in SPECT allows to restrict the detected signal to the spectral region surrounding the full-energy deposition, this energy range is also referred to as the "detection window".

\par 

\begin{figure}[htbp!]
\centering
\includegraphics[height = 6cm]{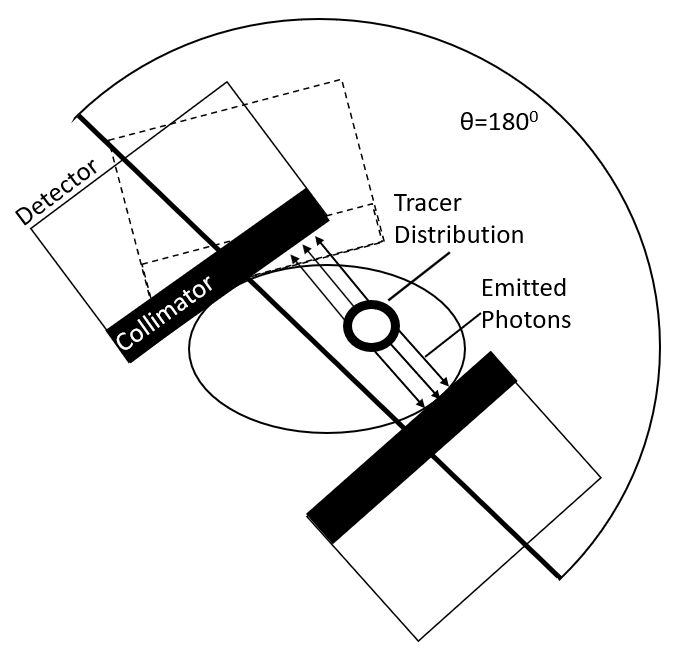}
\caption{Schematics of the SPECT imaging} 
\label{f:spect}
\end{figure}

Figure \ref{f:spect} shows the basic design of a SPECT detection head. The two detector arrays surrounding the object are used to measure the emitted photons, and each array is formed by tens of parallel, collimated, small detectors. This design is called "Dual-head SPECT". The two detector arrays, i.e., "Heads", rotate during the measurement to acquire $180^o$ or $360^o$ projections.  The collimator allows only the photons impinging on the front face of the detector be measured.

Thanks to the collimation, once a detector measures a photon, it is assumed that the radioactive tracer is located along the line perpendicular to the detector surface. At each angle, the total measured counts by a detector pixel represent the integral of the source distributions along the line, and it is called "Projection data". When rotating the detector over $180^o$ or $360^o$, we can obtain the complete projection data and use it to create final source distribution images, by reconstruction methods like the Filtered Backprojection method \cite{Ramachandran1971}, which will be briefly introduced in the following section. 

High sensitivity, low cost, relative availability of radiotracers, and compatibility with cardiac implantable electronic devices made SPECT very popular for functional myocardial imaging. However, it also has some fundamental limitations, for example the scattering effect, which can degrade the image quality. When a photon is emitted from the tracer, it can undergo one or more scattering interactions during its path. Scattering reactions in general produce photons with a lower energy and different direction, compared to the original photon.

If the scattered photon is detected and its energy is still within the detection window, the estimated source location is actually the location where the scattering happens, instead of the true tracer position. This phenomenon produces image artifacts. It is also challenging to mitigate this effect, since the scattering process not only depends on the tracer depth and distance, but also on the size, shape, composition and uniformity of the patient.

\section{Deterministic methods for scattering correction}

In order to correct for the scattering effect, several deterministic methods have been developed. These methods can be classified into two categories: methods based on the estimate and subtraction of the scattering component and methods that attempt to locate the scattered photon emission point \cite{Frey1993}. The subtraction methods aim at calculating the number of scattered photons within the photo-peak measurement window and subtract them from the photo-peak integral. One example of this approach is the use of a secondary energy window to estimate the scattering contribution \cite{Jaszczak1985}.\par

In 1984, R. J. Jaszczak and colleagues \cite{Jaszczak1984} studied the scattering effect of Tc-99m photon sources inside water-filled phantoms. As expected, the photo-peak window for Tc-99m is around 140 keV (Figure \ref{f:Tc_Tl_spectrum}). Jaszczak then used an image reconstructed using counts within a low-energy pulse-height window (92-125 keV) to correct the image based on the counts within the photo-peak window (127-153 keV). This compensation method is detailed in Equation \ref{eqn:Jaszczak1984}.
\begin{equation}
    f(x,y)=f_1(x,y)-k\cdot f_2(x,y)
    \label{eqn:Jaszczak1984}
\end{equation}
$f_1(x,y)$ is the image reconstructed using the counts in the photo-peak window, $f_2(x,y)$ is the image reconstructed using the counts in the spectral region below the detection window and $k$ is a scaling factor, experimentally determined. Using this correction method, the images of Tc-99m-filled spheres showed a better contrast within the water-filled phantoms than uncorrected images, and both qualitative and the quantitative improvements are achieved for reconstructed images.\par
\begin{figure}[htbp!]
\centering
\includegraphics[height = 4cm]{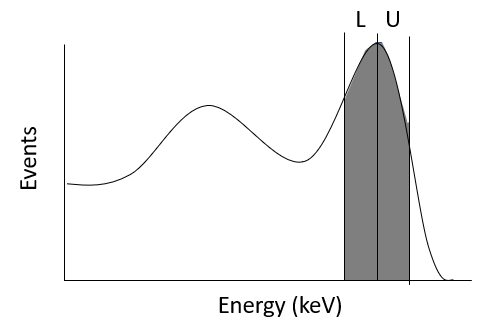}
\caption{Dual-photopeak windows in Tc-99m spectrum }
\label{f:Graph}
\end{figure}

M. A. King and colleagues also developed a subtraction method using a dual-photopeak window to correct for the scattering effect \cite{King1992}. As Figure \ref{f:Graph} shows, M. A. King divided the photo-peak window into two non-overlapping regions. 

Based on the hypothesis that more scattered photons exist in the low-energy side of the photopeak, than in the high-energy side, King developed a regression relation between the ratio of counts within two windows and the scatter fraction of the counts in the  photo-peak window, as in Equation \ref{eqn:1992_king}.

\begin{equation}
    SF=A\cdot (R_S)^B+C
    \label{eqn:1992_king}
\end{equation}

 $SF$ is the scattering fraction, $R_S$ is the count ratio of two windows, and A, B, and C are experimentally determined coefficients. By using this equation, the scattering fraction can be calculated and the scattered photons can be subtracted from the photo-peak counts.\par
 Besides these approaches, a subtraction method based on a deconvolution technique is also introduced. This method treats the scattered photons as the convolution of non-scattered image projection data and a scattering amplitude function, as shown in Equation \ref{equ:Axe_deconvolution}. The deconvolution of this scattering amplitude function from the total project data can reveal the scattered component, and the compensation can be achieved by subtracting of such component from the acquired image. 
 \begin{equation}
     S(x)=\int_{-D}^{D}P(\tau)\times F(x-\tau)d\tau
     \label{equ:Axe_deconvolution}
 \end{equation}
 In Equation \ref{equ:Axe_deconvolution}, x is the position in projection data and $S(x)$ is the scattering component at position x. $P(\tau)$ is the measured projection data, $2D$ is the image range, and $F(x-\tau)$ represents the scattering amplitude at a distance $|x-\tau|$ from the source. B. Axelsson \cite{Axelsson1984} introduced this equation and calculated the function $F(x-\tau)$ through the measurement of a line Tc-99m source in a water bath. Then this scattering amplitude function is used to estimate the scattering components and subtract it from images of a Tc-99m source in a water-filled cylinder phantom.
Besides B. Axelsson, C. E. Floyd \cite{Floyd1985} has also introduced a similar deconvolution method.\par
The methods that estimate and subtract the scattering component are typically time-efficient and relatively simple to implement, but have some unavoidable limitations. For example, the estimated scattering components are normally noisy and by subtracting them from projection data, the noise of the corrected image will increase. \par
The other scattering-correction category, which attempts to find the true origin of scattered photons, tends to be more accurate, compared to both cumulative spectrum-based correction techniques and deconvolution methods. 
One example of these methods uses a restoration filters to correct for the scattering component. For SPECT image reconstruction, Filtered Backprojection is a frequently-used technique. The FBP method is an image reconstruction method that obtains the projection data by integrating the source counts, as attenuated by the imaged subject, along a single line. FBP then applies the filtered inverse Radon transform to reconstruct images from projected counts.
During the backprojection reconstruction, restoration filters are applied to the frequency domain to correct for the inherent blurring in projection and reduce the high-frequency noise. M. A. King used a Metz frequency filter to correct for the image degradation and mitigate the scattering effect \cite{M.A.King1991, King1988}. 
Metz filters are restoration filters, which are image-dependent and minimize the normalized mean square error (NMSE). Other than the Filter-based method, iterative reconstruction methods calculate the scatter response function (SRF)  and iteratively find the distributions of the scattered photons. The scatter response function can be expressed by Equation  \ref{eqn: scatter response function}:
 \begin{equation}
     A_j=F_{ij} V_i
     \label{eqn: scatter response function}
 \end{equation}
$V_i$ is the source voxel, $A_j$ is the detector pixel, and $F_{ij}$ is the scatter response function, which describes the probability of the photons in the source voxel $V_i$ to undergo scattering interactions and are measured by the detector pixel $A_j$. If the SRF of the imaged object is known, the scattering distribution can be calculated and the image degradation can be corrected for. Based on this idea, E. C. Frey and colleagues developed a slab derived scatter estimation (SDSE) method to calculate the scatter response function of various objects and incorporate it into iterative reconstruction techniques to compensate for scattering effects \cite{Frey1993}. 
 However, this SDSE method only works well for uniform scattering medium. In nonuniform medium, the element compensation, density and distribution can make SRF calculation become very complex. 
 Moreover, Frey \cite{Frey1996} found that for SPECT with Tl-201 agents, the SDSE method did not perform as effectively as for SPECT with Tc-99m agents because of its low energy window. So Frey developed an alternative method, the effective source scatter estimation (ESSE), for the Tl-201 agents. Besides using SRF to correct for scatter, F. J. Beekman and colleagues developed another compensation method \cite{Beekman1993, Beekman1996}. They developed an analytical expression of the point spread function, which describes the photon count density distribution from a point source, and included an object shape-dependent scatter term and applied it in the iteration reconstruction process to correct for the scattering.\par
 The deterministic methods described above do not require long computational times for image reconstructions. However, they tend to perform better in simple uniform phantoms. When processing the correction under complicated geometries, like the thorax region, the accuracy of scatter corrections would be degraded. In this case, the Monte Carlo based correction methods will be more robust when applied to non-uniform, heterogeneous lattices, hence more appropriate to mitigate the scattering effect and improve the image quality.
 
 \section{Monte Carlo based methods for scatter correction}
 
 \subsection{Inverse Monte Carlo method}
 
 Monte Carlo simulation can track the emission and transport of single photons and simulate all the possible interactions including scattering events inside an object. By knowing this information, we can evaluate the number of scattered photons in the total detector-measured counts and their origin locations, and then perform the scattering compensation for the SPECT images through an inference approach that minimizes it.
 
 In 1986, Floyd and colleagues \cite{Floyd1986} developed an Inverse Monte Carlo (IMC), which can simultaneously compensate scattering effect while estimating the source distributions.
 
\begin{figure}[!h]
		\centering
		\includegraphics[width=.4\textwidth]{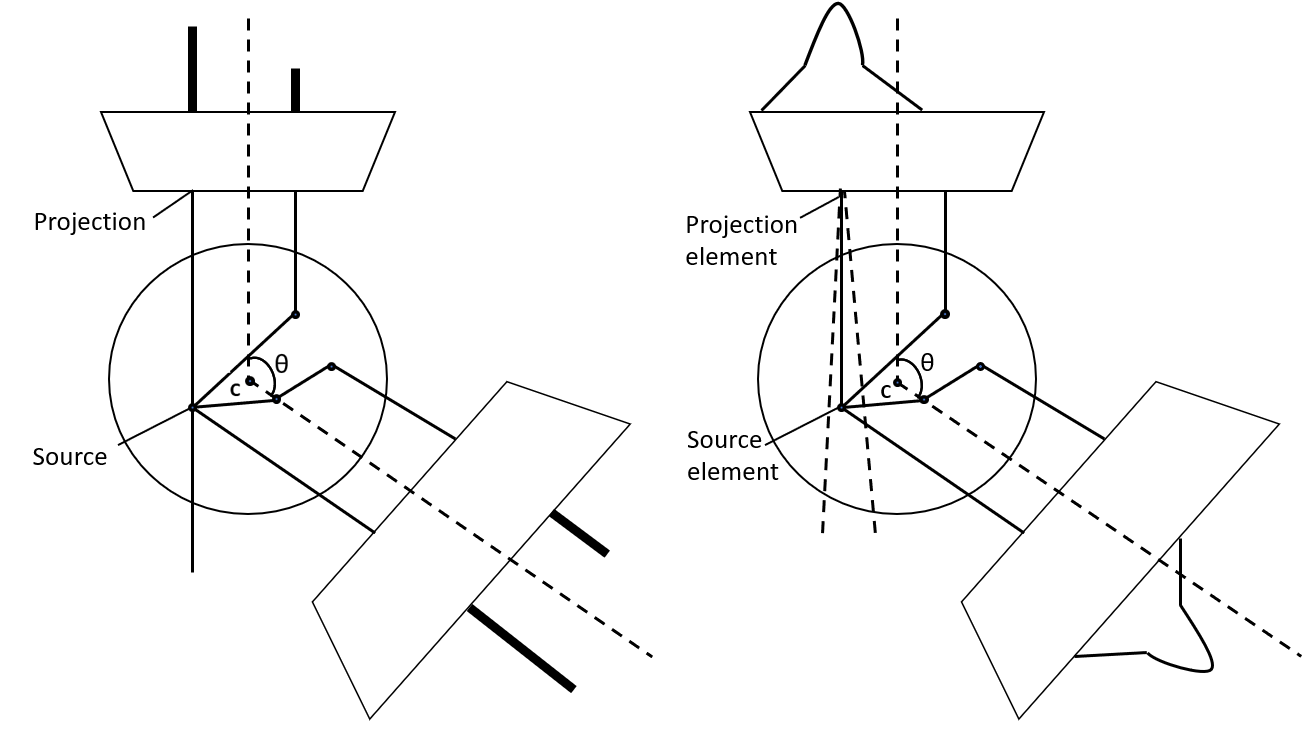}
		\caption{Schematics of SPECT acquisitions. (Left represents the filter-backprojection method. Right represents the Inverse Monte Carlo method )}
		\label{f:IMC}
\end{figure}

As Figure \ref{f:IMC} (a) shows, the FBP method treats the projection data as a perfect line integral of the sources. However, the divergence of the collimator, the attenuation and scattering of the photons can all bring unaccounted errors. For the IMC reconstruction, the measured projection data can be expressed as in Equation \ref{eqn:IMC}: 
 
 \begin{equation}
  P_j=\sum_{i=1}^N T_{ij}S_i
     \label{eqn:IMC}
 \end{equation}
 
 $P_j$ is the projection counts in detector pixel j, $S_i$ is the source voxel i, N is the number of source voxels, and $T_{ij}$ is the response matrix which represent the probability that the photons in voxel i can be detected in pixel j. The reconstruction procedure of IMC consists in first using the Monte Carlo method to calculate the response matrix $T_{ij}$, then use the measured projection data $P_j$ to solve the inverse problem and get the source distribution $S_i$. In this method, the photon attenuation and scattering effects are accounted for in the response matrix $T_{ij}$, so they can be simultaneously corrected for during the reconstruction. However, due to the high statistical uncertainty, the exact solution for Equation \ref{eqn:IMC} does not exist. 
 Therefore, the Maximum Likelihood method through the iterative Estimation-Maximization (MLEM) \cite{Shepp1982, Lange1984,Dempster1977} is used to find the solution with the minimum variance. C. E. Floyd \cite{Floyd1987} incorporated the iterative estimation maximization (MLEM) algorithm into IMC to solve for the source distribution. The applied MLEM algorithm can be expressed as in Equation \ref{eqn: MLEM}.
 
 \begin{equation}
     S_i^{k+1}=\frac{1}{\sum_{j=1}^{N_p}T_{ij}}S_i^k \sum_{m=1}^{N_p}\frac{P_m T_{im}}{\sum_{l=1}^{N_s}S_l^kT_{lm}}
     \label{eqn: MLEM}
 \end{equation}
 $S, T, P$ have been introduced in Equation \ref{eqn:IMC}. $N_p$ is the number of projection elements, and $N_s$ is the number of source elements. k is the iteration number and the term $\sum_{l=1}^{N_s}S_l^kT_{lm}$ is the reprojection, or simulated projection data. This IMC reconstruction method outperforms the FBP method in terms of accuracy. However, due to the computational time limit, this IMC method can only be incorporated in 2D reconstructions. For 3D reconstruction, the simulation of the entire response matrix and the reconstruction process requires too much computational time to make it appropriate for clinical usages.\par
 
 \subsection{Variance reduction techniques}
 
\begin{figure}[htbp!]
\centering
\includegraphics[height = 6cm]{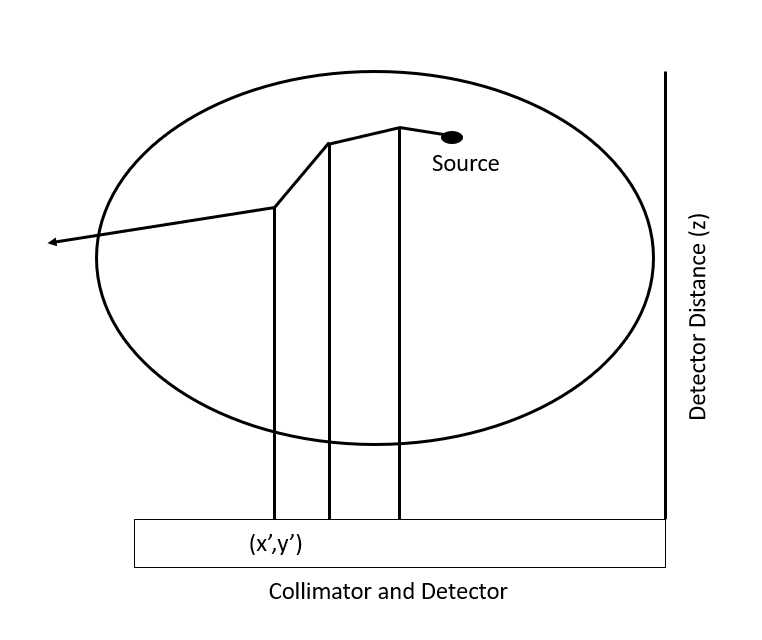}
\caption{Schematics of forced detection process}
\label{f:FD_schem}
\end{figure}

 The scatter correction method described in the previous section, which uses the general Monte Carlo algorithm to calculate response matrix, has in the long computational time its main limitation. Monte Carlo simulation consists of random samplings which will track and sample every step of emission photon's interactions, directions and energies, etc.
 The simulation process can be accurate but very time-consuming. Furthermore, during the simulation process, most of the emitted photons are attenuated by the objects and the collimators before being detected in projection bins, so they do not effectively contribute to response matrix and they are the useless photons. 
 This means that if we try to calculate a response matrix with small statistical errors, we need to simulate enough useful photons and the total number of photons which needs to be simulated could be very large. This will lead to an enormous computational time, and it is exactly the biggest setback for MC reconstruction methods. Various variance reduction techniques (VRT) have been developed to reduce the computational time of Monte Carlo simulation. Beck \cite{Beck1982} has introduced several variance reduction techniques to only keep useful photons simulated in Monte Carlo simulations. These techniques are:\\
 1) \textbf{Forced direction}\\
     For standard simulation, the initial photons are evenly emitted in 4$\pi$ directions isotropically. Photons that move away from detector surface cannot be detected and become useless photons. So we can force the initial photons only to be emitted within a solid angle $\Omega$ towards the detectors, and weight them by certain values. The weight can be expressed as the following ratio $\frac{\text{  $\Omega$} }{\text{ $4\pi$}}$. After the forced direction, all photons will be emitted towards the detector and have a smaller weight. This will have the same simulation result but lower associated statistical error due to the higher number of useful photons that are simulated, and can reduce the required computational time.\\  
2) \textbf{Forced interaction}\\
     During Monte Carlo simulations, various physical processes will be simulated, like photon scattering process, photo-electric process, etc. However, if an initial photon goes through a photo-electric interaction and is absorbed by the imaging object or SPECT's collimators, it will not contribute to the final response matrix result. So we can force the photons only go through Compton scatterings at every interaction point, and weight the scattered photons by a certain values. This value can be expressed as the following ratio $\frac{\text{Probability of Compton scattering occurs} }{\text{Probability of any interaction occurs }}$. Through this process, we will not simulate any absorbed photons that do not contribute to the image, so we can reduce the computational time but keep the same final simulation result.\\
 3)  \textbf{Forced detection}\\
Forced detection consists in forcing the photons to be detected at each interaction location, and we weight it by certain values. This process is outlined in Figure\ref{f:FD_schem}.

As Figure~\ref{f:FD_schem} shows, an initial photon is emitted at the source location and goes through several interactions in its path. At each interaction point, the initial photon will duplicate into two photons, the initial one and a copied one. The initial photon will keep following its original physical process in the object and the copied photon will be forced to be detected by the detector. After being detected, this photon will be multiplied by a weight factor. The weight factor is the product of the probability that it does not get absorbed at the interaction location, the probability that it has a certain moving direction perpendicular to the detector surface after scattering, and the probability that it is not absorbed by the medium and collimator during its path toward detector.\par

\begin{figure}[htbp!]
\centering
\includegraphics[height = 6cm]{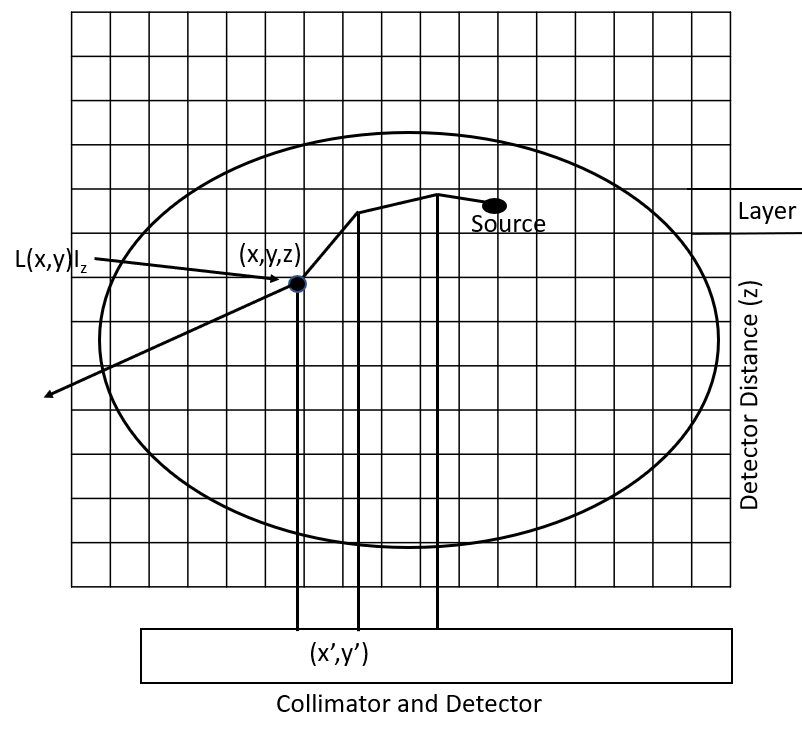}
\caption{Schematics of convolution-based forced detection process}
\label{f:CFD_schem}
\end{figure}

These three variance reduction techniques are frequently used in Monte Carlo simulations to deal with rare events and improve the simulation efficiency. Since they apply various weights to bias the simulation, they are usually referred to as the Biased Monte Carlo methods. However, it still could take hours to generate the response matrix through Monte Carlo simulation, even when the variance reduction techniques are applied. In order to further reduce the computational time, H. de. Jong \cite{DeJong2001} developed a convolution-based force detection (CFD) technique, which is one or two orders of magnitude faster than the standard forced detection technique. This CFD method is shown in Figure \ref{f:CFD_schem}.

Photons undergo an analogous process to the standard force detection technique, in which a second photon, equal to the first one, is created, weighted and detected. The only difference is that for the standard FD, the copied photon is forced to travel along a path perpendicular to the detector surface and be detected with a certain weight. However for CFD, when an interaction happens, the weight of the copied photon is stored in a sub-projection map with different layer bins. After all initial photons are simulated, the layer bins with the weight values will be convolved with a distance-dependent detector response function $PSF(x,y)|_Z$ to create the final projection data. 
Through this procedure, it is not necessary to continue sampling the copied photons until they are detected, and the convolution process can greatly decrease the computational time. In order to validate the method, H. de. Jong used standard FD and CFD to generate the scatter projections of Tc-99 point sources and extended sources in the Mathematical Cardiac Torso, and computed their Normalized Mean Square Errors (NMSE) with reference projections. The results show that CFD has a NMSE over ten times lower than FD when simulating the same number of photon histories. 
Since CFD needs to simulate fewer photons to have a statistical error comparable to standard FD, it can greatly reduce the MC computational time for scatter correction.\par

\subsection{3-D SPECT reconstructions with MC based scatter correction}

As mentioned above, the early MC-based scatter compensation method developed by Floyd \cite{Floyd1986} cannot be applied on 3-D reconstructions due to the computational time limitations. However, with the developments of several computational time reduction techniques, F. J. Beekman presented an efficient fully 3-D iterative reconstruction method for SPECT, with MC-based scattering compensations \cite{Beekman2002} in 2001. Compared to Floyd's method, Beekman has made three major improvements to significantly reduce the computational time and make 3-D reconstruction efficient.\par
The first improvement is, instead of using the MLEM method, Beekman applied Ordered Subsets Expectation-Maximization (OSEM) \cite{Hudson1994,Kamphuis1996} with dual matrix for reconstructions. The difference between them is that OSEM will divide the projection data into different subsets. As Equation \ref{eqn: MLEM} shows, the MLEM method estimates the source distribution $S_i$ by iteratively calculating response matrix $T_{ij}$ and projection data $P_j$. The response matrix is achieved by Monte Carlo simulation, which includes attenuation, detector blurring and scattering effect. Because the scatter part is the most complicated part and directly leads to a huge size of the response matrix, it confines this MLEM method in 2-D reconstructions. 
Another MLEM feature negatively affects computation time. For this reconstruction method, the projection data is divided into multiple subsets. 
Through the MLEM method one subset at a time for each iteration is processed. In order to accelerate the reconstruction, Beekman introduced the Dual Matrix Ordered Subsets (DM-OS) method which used a dual matrix on OSEM method, as expressed by Equation \ref{eqn: DMOS}. The use of dual matrix will reduce the matrix size, and the use of ordered subsets expectation maximization will process all subsets of projection data one time, instead of one subset at a time to be more efficient.
\begin{equation}
   \lambda_i^{n+1}(k)=\frac{1}{\sum_{j=1}^{S_p(n)}b_{ij}}\lambda_i^n(k) \sum_{j=1}^{S_p(n)}\frac{P_j b_{ij}}{\sum_{i=1}^N\lambda_i^n(k)c_{ij}}
     \label{eqn: DMOS}
\end{equation}
$\lambda^{n+1}(k)$ is the updated activity after processing subset $n$, $S_p(n)$ are the projection angles of subset $n$ and k is the iteration number, and N is the number of source voxels. Compared to the MLEM method, DMOS method uses two matrices $b_{ij}$ and $c_{ij}$. $c_{ij}$ is the reprojection matrix including attenuation, detector blurring and scatter distributions, and $b_{ij}$ is a much smaller back-projection matrix which only considers the attenuation and detector blurring. Because of the use of $b_{ij}$, DMOS becomes a simplified algorithm for scatter correction, which implements the reconstruction process up to two orders of magnitude faster than MLEM, and can be implemented to 3-D reconstructions.\par
The second improvement of the efficient fully 3-D iterative reconstruction method is to use CFD as variance reduction techniques, rather than standard FD to accelerate Monte Carlo simulations. The CFD method was described in the previous section and proved to be tens of times faster than standard FD.\par

The last improvement consists in reusing the calculated photon tracks (RPT) from previous iterations to decrease the reconstruction computational time. As Equation \ref{eqn: DMOS} showed, at each iteration step, in order to get the update source distribution $\lambda ^{n}(k+1)$, we need to obtain the reprojection data of the whole source distribution $\sum_{i=1}^N \lambda_i^n(k)c_{ij}$ through simulation. Since the photon transport simulation of the entire source distribution is time consuming, it requires a large amount of computational time for the reconstruction process. So the RPT technique provides another way to calculate reprojection data $\sum_{i=1}^N \lambda_i^n(k)c_{ij}$. At iteration (k-1) and (k), we will obtain the source distributions $\lambda^n(k-1)$ and $\lambda^n(k)$ and a simulated projection data $\sum_{i=1}^N \lambda_i^n(k-1)c_{ij}$. Instead of getting $\sum_{i=1}^N \lambda_i^n(k)c_{ij}$ through simulating the whole transport process for a distributed source $\lambda^n(k)$, the RPT method takes subtraction $\lambda^n(k-1)$ from $\lambda^n(k)$ to get the source distribution difference $d^n(k)=\lambda^n(k)-\lambda^n(k-1)$, and only simulates the projection data of the source distribution difference $\sum_{i=1}^N d_i^n(k)c_{ij}$, and then obtain the updated simulated projection data $\sum_{i=1}^N \lambda_i^n(k)c_{ij}= \sum_{i=1}^N d_i^n(k)c_{ij}+\sum_{i=1}^N \lambda_i^n(k-1)c_{ij}$. Through this method, at each iteration process, we only need to simulate a small fraction of sources $d^n(k)$ instead of whole distributions $\lambda^n(k)$ to obtain the updated reprojection data and it can greatly reduce the computational times.\par
By applying these techniques, Beekman developed an efficient 3-D reconstruction method with MC-based scatter compensations. Beekman and colleagues tested the algorithm by reconstructing Tc-99m sources in a water-filled phantom, and the result shows that this MC-based stochastic method outperforms the method with advanced analytical scatter model, and it is relatively efficient. This reconstruction takes approximately 20 minutes for one iteration with $10^7$ photons simulated by a single 1.4GHz processor, and almost all the computational time is used to simulate reprojections. Beekman also indicated that this 3-D reconstruction method may still be too slow for clinical usages. However, most of the reconstruction time is from the Monte Carlo simulation part and it can be accelerated by using multiple processors. So as the computational abilities of commercial computers developed rapidly, this 3-D reconstruction method can be practical for clinical routine applications.\par
Since the total computational time strongly depends on the number of simulated photons at each subset, choosing appropriate photon numbers is important for efficient reconstructions. In 2005, T. C. de Wit and colleagues \cite{DeWit2005} tested Beekman's 3-D Monte Carlo based reconstruction method with different numbers of simulated photons and tried to find the minimum photon number which is sufficient to have high quality images. T. C. de Wit used Beekman's 3-D reconstruction method to reconstruct the Tc-99m perfusion images in an anthropomorphic thorax phantom and performed it with $10^4$, $10^5$, $10^6$ and $10^7$ photons per subset in MC simulations. Then they compared the final estimated source distributions in five myocardial regions. The result shows that $10^5$ photons estimated the source distribution which has less than $1\%$ deviation comparing to $10^7$ photons and no marked improvement can be obtained with more than $10^5$ photons estimated per subset. Even though the optimized photon numbers should depend on different phantoms and source distributions, $10^5$ photons per subset should be sufficient for most myocardial SPECT imaging cases. Based on Beekman's computational time study \cite{Beekman2002}, the 3-D reconstruction time with $10^5$ photons takes less than 2 minutes per iteration by one 1.4GHz processor. This fast speed can make it practical in real clinical applications.\par

\subsection{Phantom study of 3-D reconstruction method with MC based scatter compensations}

\begin{figure}[htbp!]
\centering
\includegraphics[height = 6cm]{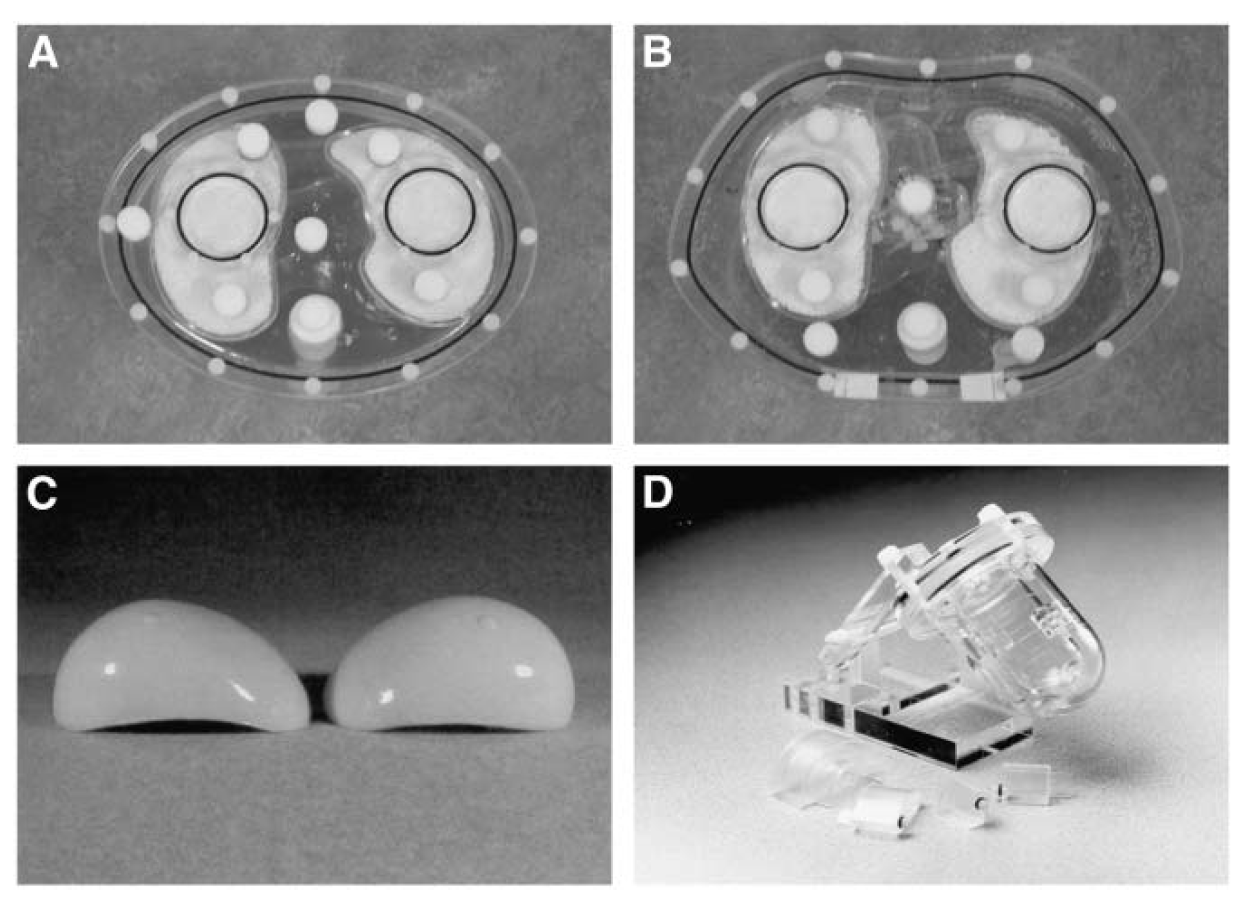}
\caption{Thorax phantoms used by Xiao \cite{Xiao2006}. (A: small phantom. B: large phantom. C:breast attachments. D: myocardial insert with solid defect set.)}
\label{f:Xiao_2006_phantom}
\end{figure}

Since the accuracy of scatter compensations strongly depends on the complexity of the imaged object, the 3-D reconstruction method proposed by Beekman still needs to be further validated under realistic anatomic configurations, especially complex configurations like real human thorax. In 2006, J. Xiao \cite{Xiao2006} tested the scatter correction ability of this method for Tc-99m cardiac perfusion SPECT on four clinical realistic phantoms, a large anthropomorphic thorax with and without breast model and a small anthropomorphic thorax with and without breast model. The phantoms are showed in Figure \ref{f:Xiao_2006_phantom}.
Two solid defects were put in the anterior and inferior walls of the cardiac insert. The Tc-99m pertechnetate solution filled the ventricular wall of the cardiac insert and a dual head Philips camera was used to detect the emitted photons.

\begin{figure}[htbp!]
\centering
\includegraphics[height = 10cm]{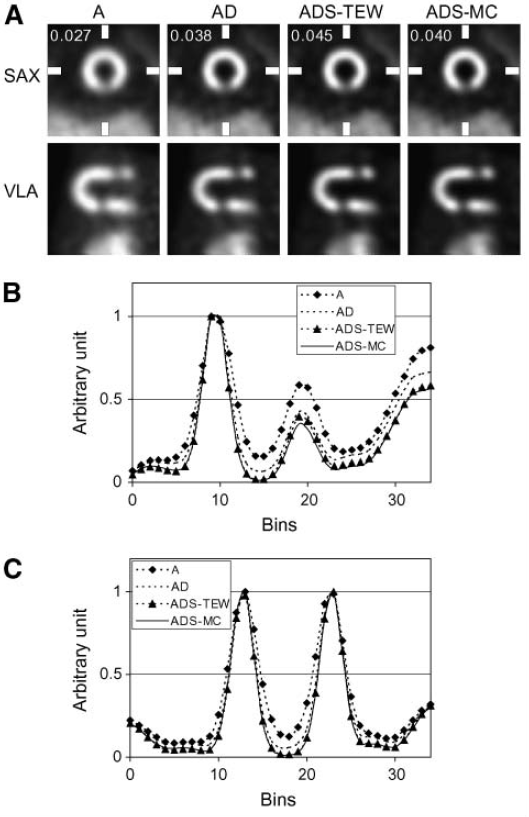}
\caption{Large thorax with breast reconstruction results\cite{Xiao2006}. (A: Top row: short-axis views of reconstructions. Bottom row: vertical long-axis reconstructions. B and C: Vertical and horizontal profiles through inferior detect in short-axis view imgaes)}
\label{f:Xiao_2006_result1}
\end{figure}

\begin{figure}[htbp!]
\centering
\includegraphics[height = 9cm]{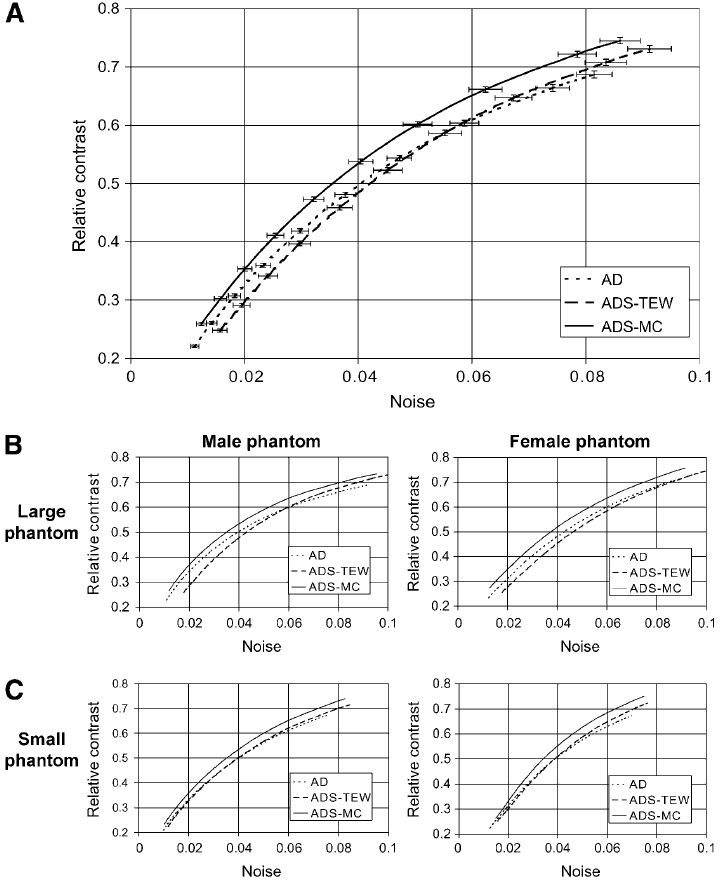}
\caption{Defect contrast as a function of noise in myocardium \cite{Xiao2006}. (A: Averaged value through four phantoms. B and C: results of individual phantom test ))}
\label{f:Xiao_2006_result2}
\end{figure}

The images were then reconstructed by ordererd-subset expectation maximization (OS-SM) algorithm through four different methods: the first method only compensates attenuation effect, the second method only compensates attenuation and detection blurring, the third method compensate attenuation and detector blurring, also use Beekman's Monte Carlo method to correct scattering, and the last method is the same as the previous one, except using multiple-window detection model to correct scattering. And finally, the reconstructed images were compared with respect to their contrasts, noises, scatter compensation abilities. 
The reconstruction results are showed in Figure \ref{f:Xiao_2006_result1} and \ref{f:Xiao_2006_result2}.

And the result shows the reconstruction method with the MC based scatter correction model creates the lowest noise when all images have the same contrast level, and it has the best scatter compensation ability which outperforms the multiple-window detection model to correct scattering. The only setback is that the MC based compensation method has the longest reconstruction time (9 min). However, it can be further reduced by using multiple processors.\par
\begin{figure}[htbp!]
\centering
\includegraphics[height = 7cm]{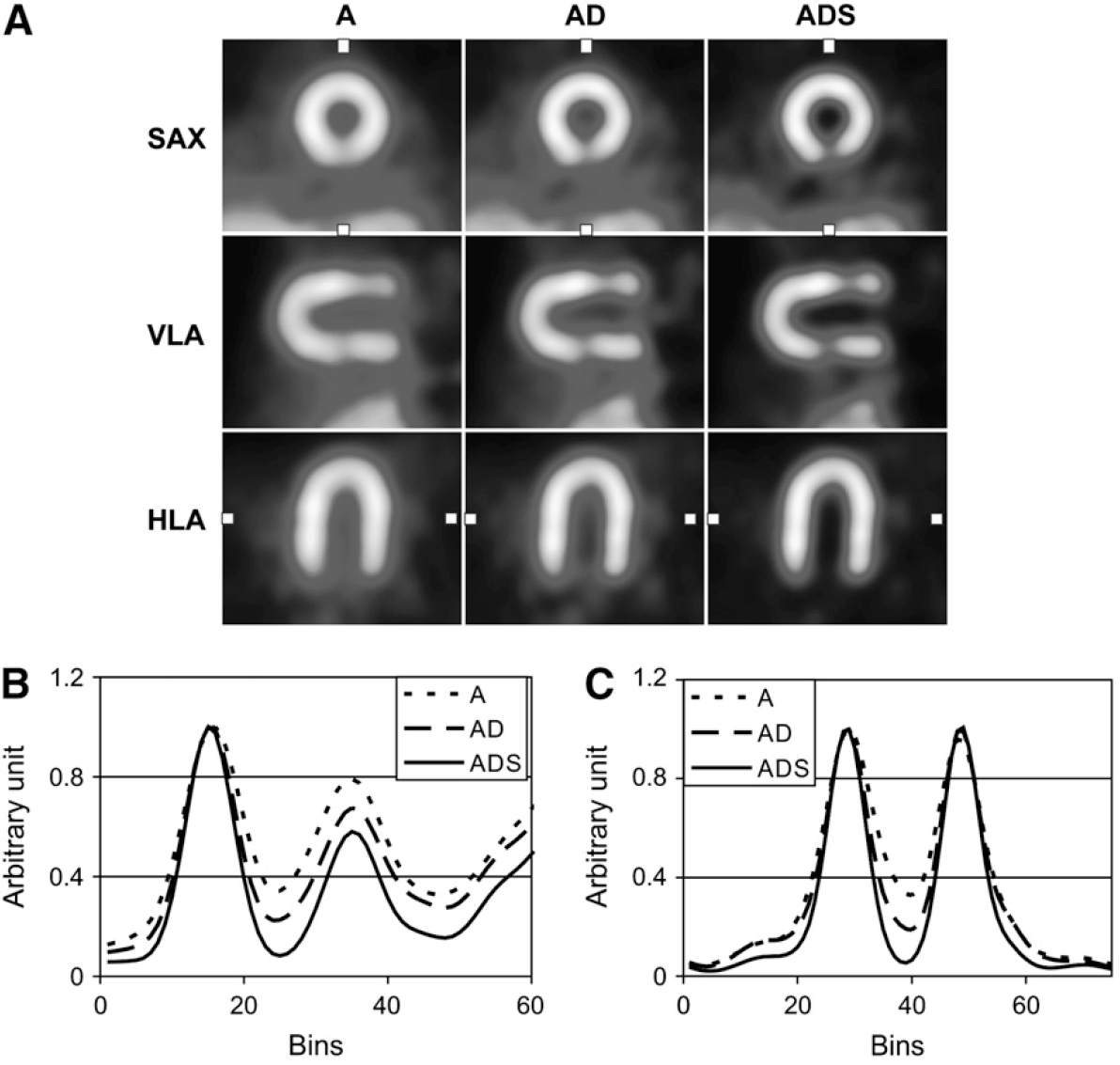}
\caption{Large thorax with breast reconstruction results \cite{Xiao2007}. (A: Top row: short-axis views of reconstructions. Middle row: vertical long-axis reconstructions. Bottom row: horizontal long-axis images B and C: Vertical and horizontal profiles through inferior detect in short-axis view images)}
\label{f:Xiao_2007_result1}
\end{figure}
\begin{figure}[htbp!]
\centering
\includegraphics[height = 7 cm]{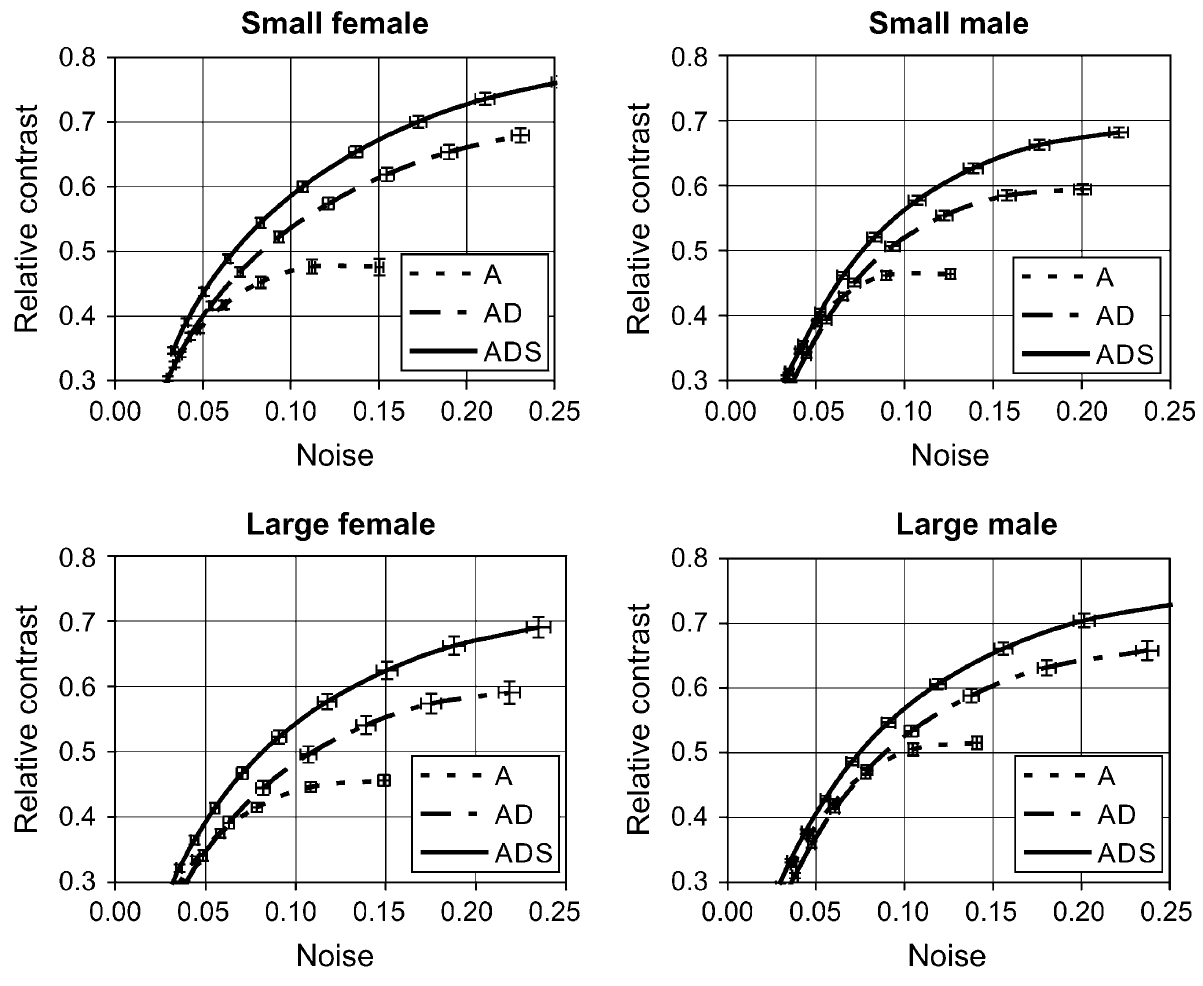}
\caption{ Defect contrast as a function of noise in myocardium for each individual phantom setup \cite{Xiao2007}}
\label{f:Xiao_2007_result2}
\end{figure}

Besides Tc-99m, another frequently-used radioisotope in SPECT cardiac imaging is Tl-201. Since the main photopeak energy of Tl-201 is lower than Tc-99m, the attenuation and scattering effects can be more severe than the Tc-99m based imaging. Moreover, the half-life of Tl-201 is longer than Tc-99m, which leads to overall lower activities in myocardial perfusion. Together, the Tl-201 based cardiac imaging has a higher noise level and the image quality is more easily degraded. Despite these shortcomings of Tl-201, it is still frequently used in many cardiac imaging applications. So the scatter correction of Tl-201 becomes significant in clinical applications. However, due to the complex spectrum of Tl-201, many scatter correction methods like the multiple-window method become inappropriate. In 2007, J. Xiao \cite{Xiao2007} tried to test the feasibility of Beekman's 3-D MC based scatter correction method for Tl-201 scatter corrections to improve the quality of cardiac perfusion images. Similar to J. Xiao's Tc-99m feasibility test, this experiment also used four clinically realistic phantom to represent cardiac environments: a large anthropomorphic thorax with and without breast, and a small anthropomorphic thorax with and without breast. Two solid defects were added to test the reconstruction performance. The phantom setup are the same as Figure \ref{f:Xiao_2006_phantom}. Both the detector counts within the low-energy photopeak window (72 keV) and high-energy photopeak window (167 keV) are measured. Measured data were reconstructed by OS-EM method with three different correction methods. The first method only uses attenuation maps to correct for the nonuniform attenuation in thorax. The second method used attenuation maps and point-spread function to correct both attenuation and detector responses. And the third method not only included the attenuation and detector response correction, but also applied the MC based scatter correction to compensate scatters. The reconstruction results are showed in Figure \ref{f:Xiao_2007_result1} and \ref{f:Xiao_2007_result2}.

Based on the final reconstruction results, the 3-D MC based scatter correction methods proved to have the best defect contrast to detect cold defects and have the noise level around 30$\%$ lower than non scatter-corrected methods. Through this experiment, J. Xiao validated the feasibility of this MC based scatter correction method in clinical applications of Tl-201 cardiac perfusion imaging.\par
\begin{figure}[htbp!]
\centering
\includegraphics[height = 6cm]{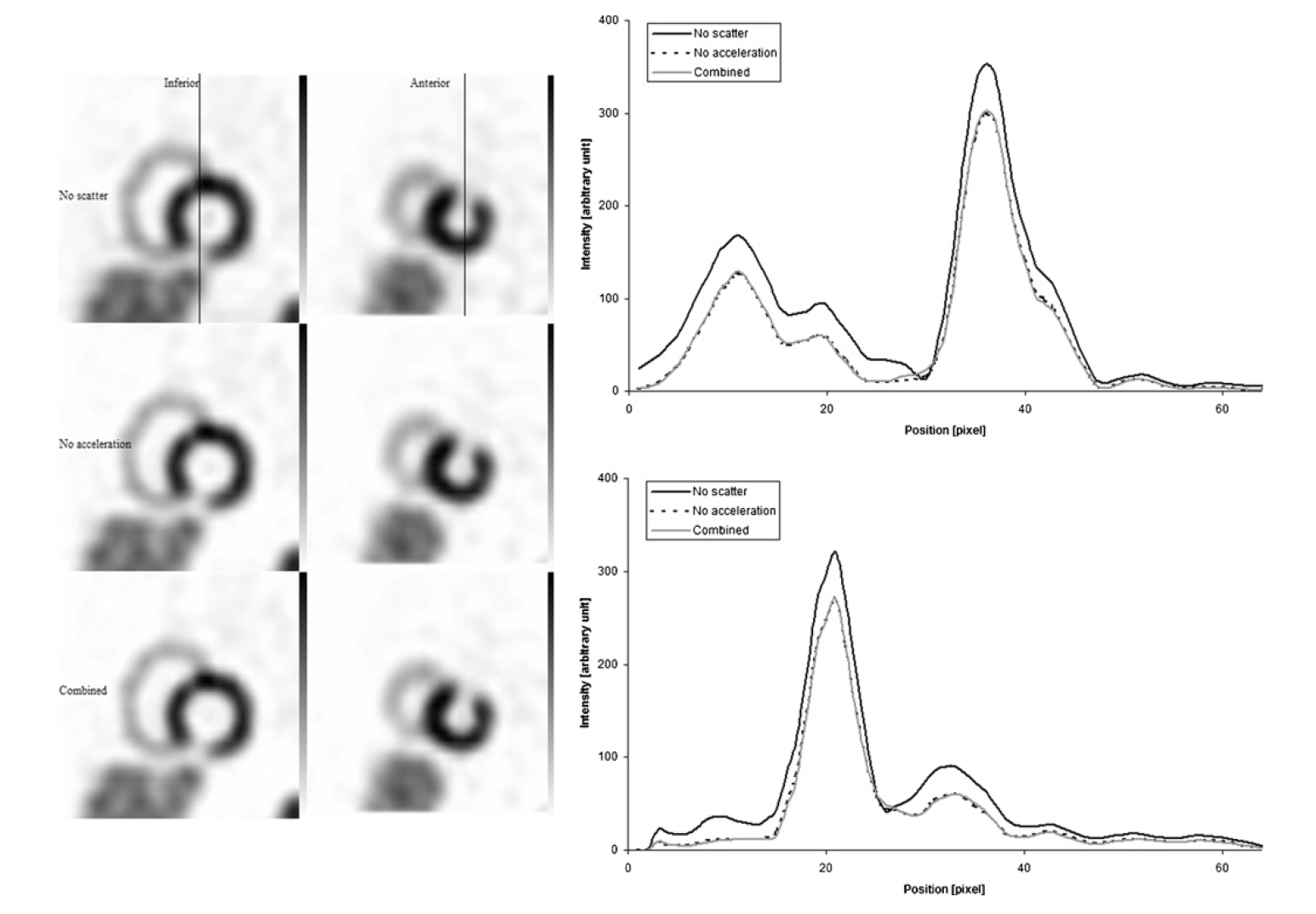}
\caption{MCAT reconstruction results\cite{Sohlberg2008}. (Left side: Short-axis slices through the heart of the phantom. Right sied: short-axis profiles, upper profile shows inferior lesion and lower profile shows anterior lesion)}
\label{f:Sohlberg_2008_result1}
\end{figure}

Since the accuracy of the MC based scatter correction has been widely proved, further improvements should aim at further accelerating the Monte Carlo simulation process and reduce the computational time. In 2008, A. Sohlberg \cite{Sohlberg2008} implemented two effective acceleration method into MC based scatter correction process to reduce the computational time, and the two methods are coarse grid scatter modelling method (CGSM) and intermittent scatter modelling. These two methods were first introduced by D. J. Kadrmas in 1998 \cite{Kadrmas1998} to accelerate the reconstruction process which models the Scatter Response Function (SRF) and projects the current image estimation to find the scatter estimations. The idea of the CGSM method is based on an observation which is that the scatter component of the projection data is normally dominated by low-frequency information. So CGSM method uses regular source voxel size to project image estimation but uses a larger source voxel size to project scatter estimation. By using the larger size of voxels, the size of the response matrix decreases and the computational time can be reduced at each iteration. Even though the scatter estimation is projected by a larger voxel size, the projected scatter data in each large pixels are expanded and interpolated into normal pixel size. The second method is referred to as intermittent scatter modelling, is based on another observation that the projected scatter estimation converges after a few iterations. So intermittent scatter modelling method only updates the scatter estimation in the very first iterations, and then keeps it constant for the following iterations to reduce the computational time. As D. J. Kadrmas showed, the usage of these two method can reconstruct images with similar qualities compared to the standard iterative reconstruction method, but can significantly reduce the computational time. Sohlberg tried to implement these two methods into MC based scatter correction method to further reduce the MC reconstruction time. Sohlberg applied the OS-EM reconstruction as showed in Equation \ref{eqn: Sohhlberg_OSEM}. 
\begin{equation}
 f_j^{new}=\frac{f_j^{old}}{\sum_{i=1}^{S_n}a_{ij}}\sum_{i=1}^{S_n} a_{ij} \frac{p_i}{a_{ik}f_k^{old}+s_i}
    \label{eqn: Sohhlberg_OSEM}
\end{equation}
 It is similar to Equation \ref{eqn:IMC} and Equation \ref{eqn: DMOS}, where $f$ is the reconstructed image, $p$ is the measured projections, $a_{ij}$ is the response matrix and $s_i$ is the scatter projections calculated by MC. A. Sohlberg incorporated the two acceleration methods, which use a larger source voxel size for MC to calculate scatter projections and only update the scatter projections in very first iteration, then keep it constant for following iterations. And finally, Sohlberg validated his method with data measured from simulated cardiac torso (MCAT) phantom with Tc-99m as the radioactive tracer and data from real clinical Tc-99m myocardial stress/rest perfusion study. Sohlberg et al. reconstructed the images by the MC based scatter compensation method with and without accelerations. The results of this approach are shown in Figure \ref{f:Sohlberg_2008_result1}.

\begin{figure}[htbp!]
\centering
\includegraphics[height = 6cm]{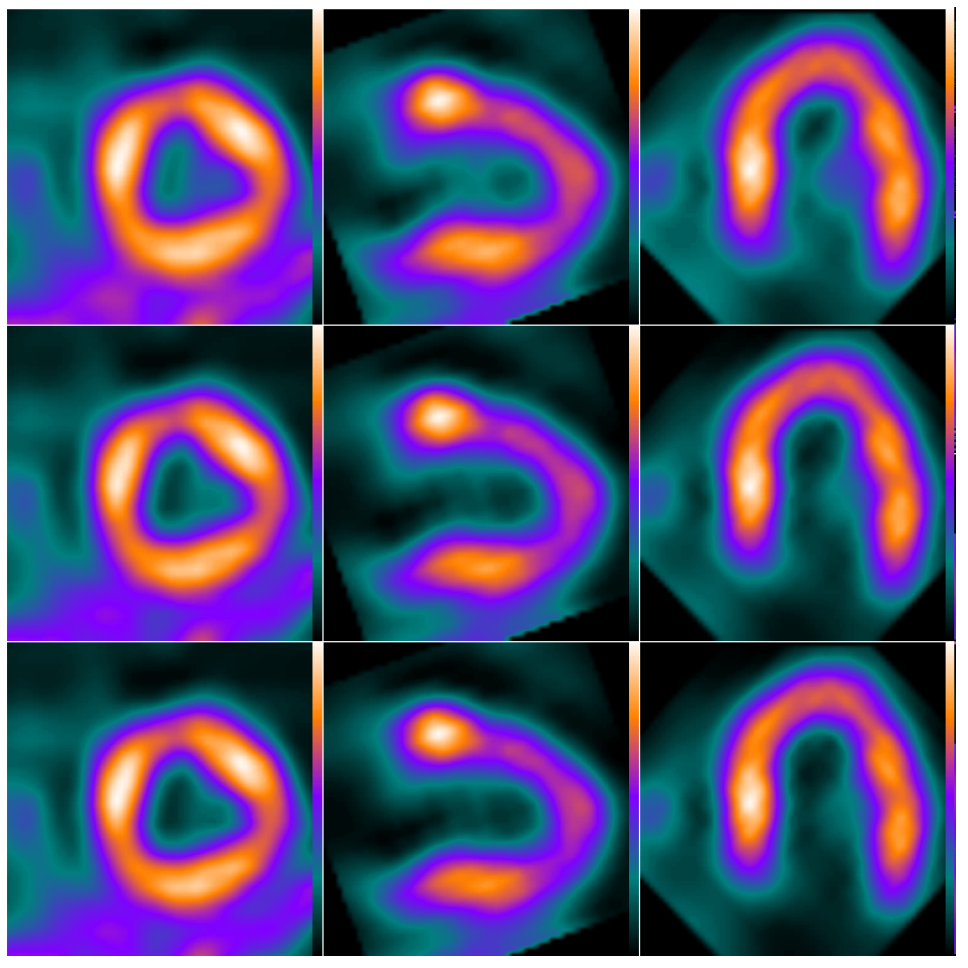}
\caption{Reconstructions of resting clinical myocardial perfusion images \cite{Sohlberg2008}. First to third columns: short-axis slices, vertical long-axis slices and horizontal long-axis slices with 1)without scatter correction, 2)MC-based scatter correction without acceleration, 3)with correction and acceleration.}
\label{f:Sohlberg_2008_result2}
\end{figure}
As the result shows, the reconstruction method accelerated by the two modelling methods can yield similar images, and it requires a computational time up to 4 times less than non-accelerated reconstruction methods. Hence these two methods are suitable for MC-based scatter correction to reduce computational time, while keeping a good image quality. \par
\begin{figure}[htbp!]
\centering
\includegraphics[height = 3cm]{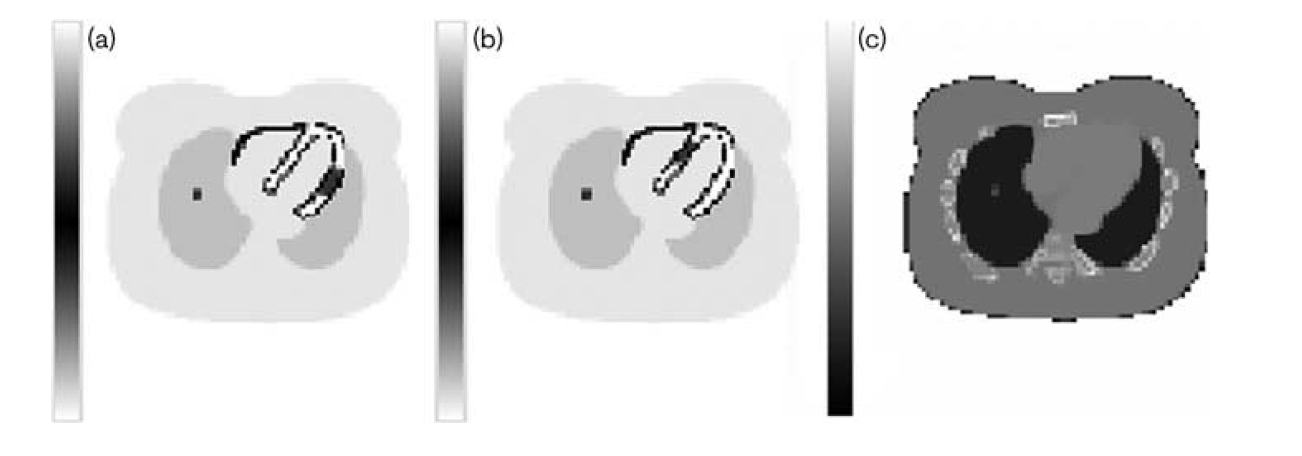}
\caption{Simulated phantom slices for the two lesion positions (a and b) and attenuation map (c) \cite{Kangasmaa2011}.}
\label{f:Kangasmaa_2011_phantom}
\end{figure}
\begin{figure}[htbp!]
\centering
\includegraphics[height = 3cm]{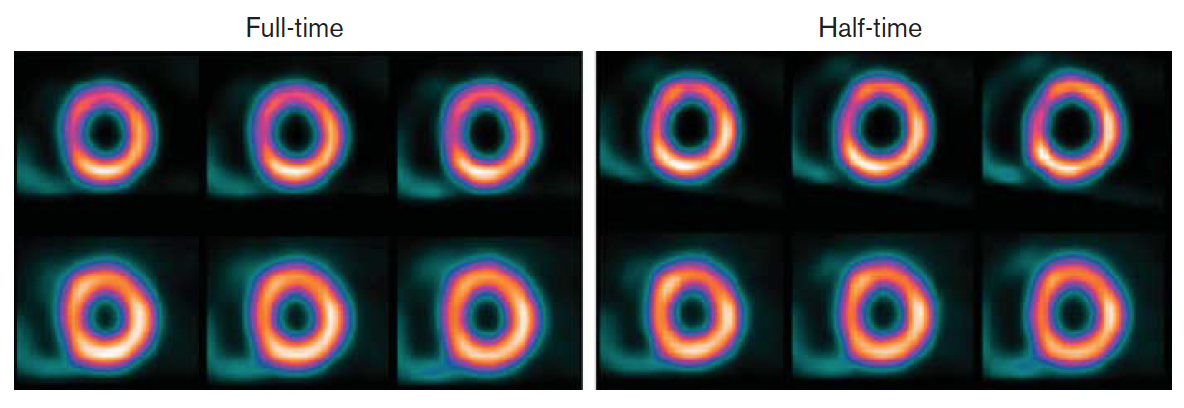}
\caption{Short-axis slices of the simulation phantom for full-time and half-time acquisition times \cite{Kangasmaa2011}. The upper shows scatter corrected slices and lower rows shows uncorrected slices}
\label{f:Kangasmaa_2011_result}
\end{figure}

In 2001, T. S. Kangasmaa \cite{Kangasmaa2011} has further tested Sohlberg's reconstruction method with MC based scatter correction, on both simulated data and realistic patient image data. Kangasmaa and colleagues generated the SPECT projection data on a simulated cardiac torso phantom which is showed in Figure \ref{f:Kangasmaa_2011_phantom} and also used cardiac perfusion images from 30 patients to test the MC based scatter correction. 
The reconstruction result is showed in Figure \ref{f:Kangasmaa_2011_result}, which indicated that compared to the non-corrected reconstruction method, the MC based correction method has a higher contrast, lower noise and better defect detection ability.
The actual patient images were reconstructed with and without scatter corrections, and the quality has been evaluated by experienced nuclear medicine experts on a scale form 1 to 5, regarding of the contrast and noise performance. The evaluation showed the corrected images had a average quality grade of 4.36, comparing to grade of uncorrected images 3.73, which indicated the feasibility of MC scatter correction method to improve qualities of realistic patient images for clinical applications.



\subsection{State-of-the-art Monte Carlo approaches}
 The computational time necessary to reconstruct high quality images remains the main setback for the application of Monte Carlo algorithms to clinical practices. The improvement of these times has remained a focus point for research conducted on MC SPECT throughout the 21st century due to the timely reactions necessitated by some patient's conditions as well as the comparatively slow reconstruction time of MC. The reduction of computation times has been the result of improvements in implementation methods alongside the growth of computer capabilities and general computing power. Recent research seems to trend towards trying to incorporate MC into clinical units by improving the computational times of established Monte Carlo methods. 
    
    
    SIMIND is an MC code that simulates SPECT systems and can be used to reconstruct images. Since the reconstruction processes are compiled in a separate program, it is referred to instead as SIMREC when used to reconstruct. One advantage of SIMREC is its ability to simulate complex emission spectra. SIMREC was evaluated in 2018 by Johan Gustafsson and colleagues for the isotopes Tc-99m, Lu-177 and I-131 using phantom tests and patient examples \cite{Gustafsson2018}. The phantom tests were performed on simple geometries. The phantoms were elliptical or spherical and filled with radioactive compounds. SIMREC performed within 7\text{\%} of the actual values on these simple geometries. The patient examples were comparisons between SIMREC reconstructions and the clinical software. The output images show visible improvement over the clinical images even without extended updates. However, real anatomy is complex and this will cause long computation times in general, upwards of 6 hours while following clinical protocol in some cases. Higher resolution images are possible through extended updates but this can take up to five times longer. This approach could still be implemented for patients whose conditions do not require timely action. Should the reconstruction speed be improved upon, this method would be exceptionally useful due to the high quality of the reconstructed images. 
    
    One possible improvement on reconstruction time is the use of a computer's graphic processing unit (GPU) to perform the computations, not the central processing unit (CPU). This idea was researched by T. Bexelius and colleagues in 2018, where XCAT phantoms and patient studies were conducted with the standard CPU approach and a modified approach to fully utilize the GPU's capabilities \cite{Bexelius2018}. This modification meant that the results were not expected to be identical because the methods were not identical. That prediction was correct, as these two approaches were found to be mathematically different, with the GPU approach being slightly worse. Using a Geforce GTX 1080Ti, a powerful GPU from 2017, for reconstruction took 3.8 seconds, which is a 24 times improvement in speed when compared to the Xeon E5-1650 v4, a six core CPU from 2016. The GPUs and CPU tested produced images that were visually indistinguishable, showing that there is no drawback for the improvement in speed. The reconstruction time for attenuation and collimator response were very short for each GPU tested. This most likely means that those reconstructions are embarrassingly parallel, as a GPU is strongest when performing the same task multiple times. However, the speed of scatter reconstruction for a GPU depends heavily on the power of the part used. Higher power GPUs spent less time on each process than lower power GPUs. It would have been beneficial to also have data from CPUs with more or less cores in order to see how the speed of each section was affected by the number of cores. The methods from this experiment produce times that are close to real time processing, which could be a great benefit to hospitals when working with all applicable patients.

\section{Discussion and Conclusions}

Monte Carlo algorithms can be used to model stochastic processes, such as radiation transport and detection. In nuclear medicine, MC algorithms have been extensively used to predict and mitigate the gamma-ray scattering component. This use is particularly suited to SPECT, where, unlike PET, detection in time coincidence cannot be implemented to filter the gamma-ray scattered component. The signal component deriving from the scattering contribution leads to an inaccurate estimate of the source location. Effective scattering correction methods have a direct impact on the source localization accuracy and therefore an improved image quality and spatial resolution of the diagnostic method. 

Early deterministic approaches performed a cumulative estimate of the signal component due to scattering, based on in-phantom calibration procedure. The main limitation of this approach is the limited applicability to geometries and nuclide energies different from the calibration scenario. In the Eighties, the introduction of MC approaches allowed an event-by-event scattering modeling, that was coupled to inverse methods for image reconstruction and filtering of the scattering contribution to the image. Application of variance reduction techniques first, and advanced computational methods later, allowed a significant reduction of computation times and extension of the method to three-dimensional geometries.
Given the intrinsic difference of test cases and image quality metrics, a direct quantitative comparison of different approaches is challenging, therefore, in this work, we have analyzed and reviewed methodological and qualitative differences. 

State-of-the-art image reconstruction based on Monte Carlo algorithms is already visually superior to the current clinical practice, even when using poly-energetic complex isotopes used in modern medicine. The use of powerful GPUs will likely yield significant improvements upon the speed of clinical image reconstruction, possibly bringing MC close to or slightly faster than the clinical approach's speed allowing for real-time MC-based scattering correction.
    
\section{Acknowledgements}
We thank Dr. T. Kangasmaa and Dr. A. Sohlberg for granting us the permission to use their artwork. The authors also acknowledge support for this work from the Nuclear Regulatory Commission Faculty Development Grant 31310019M0011.
This work was also funded in-part by the Consortium for Verification Technology under Department of Energy (DOE) National Nuclear Security Administration award number DE-NA0002534.

\medskip  

\bibliographystyle{ieeetr}

\bibliography{Reference}
    
\end{document}